\documentclass[%
 reprint,
superscriptaddress,
nofootinbib,
nobibnotes,
 amsmath,amssymb,
 aps,
prl,
]{revtex4-2}

\usepackage{multirow}
\usepackage{natbib}
\usepackage{graphicx}
\usepackage{dcolumn}
\usepackage{bm}
\usepackage{hyperref}

\newcommand{\vk}{\mathbf{k}}

\newcommand{\gs}{g_*}

\newcommand{\Lt}{\mathcal{L}_2}
\newcommand{\bPkm}{\bar P_{\rm m}(k)}

\newcommand{\hgpc}{h^{-1}{\rm Gpc}}
\newcommand{\hmsun}{h^{-1}M_\odot}
\newcommand{\hmpc}{h~{\rm Mpc}^{-1}}
\newcommand{\bht}{b_{\rm h}^{(2)}}

\newcommand{\bh}{b_{\rm h}}

\usepackage{color}
\usepackage{xcolor}
\newcommand{\Reddd}[1]{{\color{black}#1}}
\newcommand{\Redd}[1]{{\color{black}#1}}
\newcommand{\Red}[1]{{\color{black}#1}}
\definecolor{darkmagenta}{HTML}{8b008b}

\usepackage{ulem}

\def\mnras{\rm{MNRAS}}

\def\aap{\rm{A\&A}}

\def\apj{\rm{ApJ}}                 
\def\jcap{\rm{JCAP}}

\begin{document}

\preprint{}

\title{First confirmation of anisotropic \Red{halo} bias
\\
from statistically anisotropic matter distributions}


\author{Shogo~Masaki}
\email{shogo.masaki@gmail.com}
\affiliation{Department of Information Engineering and Institute for Advanced Studies in Artificial Intelligence, Chukyo University, Toyota, Aichi 470-0393, Japan}
\affiliation{Department of Physics, Nagoya University, Nagoya, Aichi 464-8602, Japan}

\author{Maresuke~Shiraishi}
\affiliation{School of General and Management Studies, Suwa University of Science, Chino, Nagano
391-0292, Japan}

\author{Takahiro~Nishimichi}
\affiliation{Department of Astrophysics and Atmospheric Sciences, Faculty of Science,
Kyoto Sangyo University, Kyoto, Kyoto 603-8555, Japan}
\affiliation{Center for Gravitational Physics and Quantum Information, Yukawa Institute for Theoretical Physics, Kyoto University, Kyoto, Kyoto 606-8502, Japan}
\affiliation{Kavli Institute for the Physics and Mathematics of the Universe (WPI), The University of Tokyo, Kashiwa, Chiba 277-8583, Japan}

\author{Teppei~Okumura}
\affiliation{Academia Sinica Institute of Astronomy and Astrophysics, AS/NTU Astronomy-Mathematics Building, No.1, Sec. 4, Roosevelt Rd, Taipei 106216, Taiwan}
\affiliation{Kavli Institute for the Physics and Mathematics of the Universe (WPI), The University of Tokyo, Kashiwa, Chiba 277-8583, Japan}

\author{Shuichiro~Yokoyama}
\affiliation{Kobayashi Maskawa Institute, Nagoya University, Nagoya, Aichi 464-8602, Japan}
\affiliation{Department of Physics, Nagoya University, Nagoya, Aichi 464-8602, Japan}
\affiliation{Kavli Institute for the Physics and Mathematics of the Universe (WPI), The University of Tokyo, Kashiwa, Chiba 277-8583, Japan}

\date{\today}

\begin{abstract}
We confirm for the first time the existence of distinctive halo bias associated with the quadrupolar type of statistical anisotropy (SA) of the linear matter density field using cosmological $N$-body simulations.
We find that the coefficient of the SA-induced bias for cluster-sized halos takes negative values and exhibits a decreasing trend with increasing halo mass.
This results in the quadrupole halo power spectra in a statistically anisotropic universe being less amplified compared to the monopole spectra.
The anisotropic feature in halo bias that we found presents a promising new tool for testing the hypothesis of a statistically anisotropic universe, with significant implications for the precise verification of anisotropic inflation scenarios and vector dark matter and dark energy models.
\end{abstract}

\maketitle


\paragraph{Introduction.---\hspace{-0.8em}}


  Isotropy is a fundamental symmetry in physics. Global isotropy, or equivalently statistical isotropy, has been regarded as an underlying conjecture in cosmology. 
  Various cosmic observations also support this symmetry,\footnote{Cosmic isotropy is, of course, locally violated as observed in the fluctuations of the cosmic microwave background, while this is not equal to the global or statistical violation we mention here.}
  although slight deviations have not been entirely ruled out.
  From a theoretical point of view, broken global isotropy, known as statistical anisotropy (SA), could indicate the presence of anisotropic sources, such as vector fields.
  Various inflationary scenarios that incorporate vector fields, motivated by magnetogenesis and axion phenomenology, have been extensively explored (see e.g., Refs.~\cite{Dimastrogiovanni:2010sm,Soda:2012zm,Maleknejad:2012fw} for review).
  There are also interesting studies on vector fields in the context of dark matter and dark energy (e.g., Refs.~\cite{BeltranJimenez:2008iye,Hambye:2008bq,Graham:2015rva,Bastero-Gil:2018uel}).
  Measuring the SA can play a crucial role in diagnosing such scenarios. 


  From both theoretical and observational sides, the quadrupolar type of SA in the primordial curvature perturbations or the linear matter density field has been studied frequently, as it is a primary feature due to vector fields.
  Its magnitude is conventionally characterized by the parameter $\gs$. 
  Observational constraints on $\gs$ have been derived from the cosmic microwave background \cite{Planck:2018jri,Planck:2019evm,Planck:2019kim} and galaxy 
  \Red{clustering} \cite{Pullen:2010zy,sugiyama18}.
  Since future galaxy surveys are expected to increase the constraining power dramatically \cite{Shiraishi:2016wec}, corresponding theoretical studies have become more important accordingly.


Recently, the impacts of SA on galaxy/halo statistics have been discussed in Ref.~\cite{shiraishi23}, highlighting the presence of a distinctive galaxy/halo bias term associated with the SA of the linear matter density field, based on a simple linear bias model.
Particularly interestingly, the bias itself becomes anisotropic.
For the quadrupolar type of SA, notably, this SA-induced bias manifests solely in the quadrupole power spectra, distinguishing it completely from the conventional linear bias.
In other words, detecting a nonzero SA-induced bias in observations would provide direct evidence of the broken global isotropy, and give a chance to test underlying cosmological scenarios, e.g., anisotropic inflation and vector dark matter and dark energy models.

To verify the existence of the bias associated with SA predicted by the simple linear bias model, in this {\it Letter}, we perform cosmological $N$-body simulations incorporating the quadrupolar SA in the linear matter density field. 
We develop three estimators for the coefficient of the SA-induced halo bias, $\bht$, which are applied to the large-scale distribution of simulated halos.
We confirm the presence of nonzero contribution from the $\bht$ term for cluster-sized halos. 
The detected $\bht$ coefficient is found to be negative, with a decreasing trend as halo mass increases,
and thus the quadrupole halo power spectra can be less amplified than the monopole power spectra
in a statistically anisotropic universe.

\paragraph{SA-induced bias. ---\hspace{-0.8em}}

Following Ref.~\cite{shiraishi23}, let us briefly review the halo bias in the statistically anisotropic universe based on the simple analytic estimation.
First, SA in the matter density fields is characterized by the Legendre polynomial $\mathcal{L}_\ell(x)$.
We focus on the quadrupolar type of SA and consider the power spectrum of the linear matter overdensity fields, $\langle\delta_{\rm m}(\vk_1) \delta_{\rm m}(\vk_2)\rangle = (2\pi)^3 \delta^{(3)}(\vk_1 + \vk_2) P_{\rm m}(\vk_1)$ with
  \begin{align}
  P_{\rm m}(\vk) &=
 \left[
    1 + \frac{2}{3} g_* \mathcal{L}_2(\mu)  \right]\bPkm ,
\label{eq:pksa}
\end{align}
where $\bar P_{\rm m}(k)$ corresponds to the isotropic component of the matter power spectrum,
$\mathcal{L}_2(\mu) \equiv \frac{1}{2} \left( 3\mu^2 - 1 \right)$ with $\mu \equiv \hat k\cdot\hat d$ and $\hat k \equiv \vk/|\vk|$, and $\hat d$ denotes the preferred direction associated with the SA.\footnote{We denote the quantities in the isotropic universe with ``$~\bar~~$'' throughout this work.}
The description for $\delta_{\rm m}(\vk)$ reproducing Eq.~\eqref{eq:pksa} reads
\begin{align}
  \delta_{\rm m}({\bf k}) 
  &=\left[1 + \frac{1}{3}g_* \mathcal{L}_2(\mu) + {\cal O}(g_*^2)\right]  \bar{\delta}_{\rm m}({\bf k}) , 
  \label{eq:mdf}
\end{align}
where $\bar{\delta}_{\rm m}(\vk)$ is the isotropic part of the matter density field obeying $\langle\bar{\delta}_{\rm m}(\vk_1) \bar{\delta}_{\rm m}(\vk_2)\rangle = (2\pi)^3 \delta^{(3)}(\vk_1 + \vk_2) \bar{P}_{\rm m}(k_1)$.
Instead of Legendre polynomial, by introducing a {\it global traceless tensor field}:
\begin{align}
  {\cal G}_{ij} &\equiv g_* \left(\hat{d}_i \hat{d}_j - \frac{1}{3} \delta_{ij}\right)~, \label{eq:Gij}
\end{align}
Eq.~\eqref{eq:pksa} can be rewritten as
\begin{align}
  P_{\rm m}(\vk) &= \left[ 1 + {\cal G}_{ij} \hat{k}_i \hat{k}_j \right] \bPkm.
\end{align}
Thus the quadrupolar SA can be interpreted as an anisotropic distortion due to the existence of the global tensor field, ${\cal G}_{ij}$.
In the following, we assume that the SA is sufficiently small, $|\gs|\ll 1$, to be consistent with observations \cite{Planck:2018jri,Planck:2019evm,Planck:2019kim,Pullen:2010zy,sugiyama18}, and we will therefore evaluate relevant quantities up to the linear order of $g_*$.

In a similar way to Refs.~\cite{mcdonald09, chan12, baldauf12, Schmidt:2015xka, Assassi:2015fma, akitsu23}, under the presence of ${\cal G}_{ij}$, we expand the halo overdensity field $\delta_{\rm h}$ with $\delta_{\rm m}$ and a traceless tidal field $K_{ij} \equiv \left( \frac{\partial_i \partial_j}{\partial^2} - \frac{1}{3} \delta_{ij} \right) \delta_{\rm m}$, leading to
\begin{align}
\delta_{\rm h} = \bh \delta_{\rm m} + \frac{1}{2} \bht {\cal G}_{ij} K_{ij} + {\cal O}(\delta_{\rm m}^2, \delta_{\rm m} K, K^2) .
\end{align}
Here, the coefficients $\bh$ and $\bht$ represent linear responses on $\delta_{\rm m}$ and $K_{ij}$, respectively. Note that, in isotropic universe models, i.e., for ${\cal G}_{ij} = 0$, the second term vanishes and hence any contribution due to $K_{ij}$ appears at higher order \cite{Schmidt:2015xka}.
At linear order of $\delta_{\rm m}$, its Fourier counterpart reads
\begin{align}
  \delta_{\rm h}({\bf k}) &= \left[\bh + \frac{1}{2} \bht {\cal G}_{ij} \hat{k}_i \hat{k}_j  
  \right]  \delta_{\rm m}({\bf k}) \nonumber \\
  &= \left[\bh +  \frac{1}{3} \bht g_* \mathcal{L}_2(\mu) \right] \delta_{\rm m}({\bf k})
  ,
  \label{eq:hndf}
\end{align}
and this equation implies that the halo bias, which is the response of the halo overdensity field to the matter overdensity field, itself is anisotropic.
Thus, Ref.~\cite{shiraishi23} pointed out for the first time that, in addition to the linear bias parameter in isotropic universe, $\bh$, a new kind of bias parameter, $\bht$, is generically introduced due to the existence of the effective global tensor field, ${\cal G}_{ij}$, that is, the SA in the matter density field.\footnote{\Red{See Ref.~\cite{akitsu23} for a similar analysis on anisotropic biases induced by gravitational waves.}}

From Eqs.~\eqref{eq:mdf} and \eqref{eq:hndf}, we can evaluate the halo autopower spectrum, $P_{\rm h}$, and the halo-matter cross-power spectrum, $P_{\rm hm}$.
By expanding these power spectra in terms of the Legendre polynomials:
\begin{align}
  P_{\rm X}(\vk)&=\sum_{\ell=0, 2} P_{\rm X,\ell}(k) \mathcal{L}_\ell(\mu)~,
\end{align}
for $\rm X \in \{ h, hm \}$, we have
\begin{align}
  \begin{split}
    P_{\rm h,0}(k) &= \bh^2 \bar{P}_{\rm m}(k) , \\
    P_{\rm h,2}(k) &= \frac{2}{3}\bh \left[ \bh+\bht\right]\gs \bar{P}_{\rm m}(k) ,  \\
    P_{\rm hm,0}(k) &= \bh \bar{P}_{\rm m}(k)  , \\
    P_{\rm hm,2}(k) &= \frac{1}{3}\left[2\bh+\bht \right] \gs \bar{P}_{\rm m}(k),
\label{eq:ABC0}
  \end{split}
\end{align}
at the leading order in the perturbative expansion. One can see that if $\bht \neq 0$, the quadrupole (anisotropic) term is biased differently from the monopole (isotropic) one.
This is an interesting prediction found in Ref.~\cite{shiraishi23}.

\paragraph{Simulations.---\hspace{-0.8em}}
To examine the presence of the $\bht$ term discussed above, we perform cosmological $N$-body simulations and investigate the halo distribution in a statistically anisotropic universe.
In our setup, aside from using the statistically anisotropic matter power spectrum given by Eq.~\eqref{eq:pksa}, we assume a standard flat $\Lambda$-cold dark matter cosmology with $\Omega_{\rm m0}=0.3156,~\Omega_{\Lambda0}=0.6844,~H_0=100h=67.27~{\rm km~s^{-1}Mpc^{-1}},~n_{\rm s}=0.9645$ and $A_{\rm s}=2.2065\times10^{-9}$ \cite{planck-collaboration:2015fj}.
We use {\sc Gadget2} \cite{gadget2} as the cosmological $N$-body solver and employ $1024^3$ simulation particles \Red{with mass of $5\times10^{12}~\hmsun$} in a box with a side length of $4~\hgpc$.
For the linear matter power spectrum given by Eq.~\eqref{eq:pksa}, we compute the isotropic part of the power spectrum, $\bar{P}_{\rm m}(k)$, at $z_{\rm ini} = 31$ using a publicly available Boltzmann solver {\sc CAMB} \cite{camb} and incorporate the SA by multiplying it the factor $[1+\frac{2}{3}\gs \Lt(\mu)]$.
We fix the preferred direction of the SA to be $\hat d=(0,~0,~1)$.
Based on the obtained matter power spectrum, the initial conditions (ICs) are generated at $z_{\rm ini}$ using the second-order Lagrangian perturbation theory (2LPT) \cite{scoccimarro98,Crocce06a,nishimichi09}.
We generate ICs using grid pre-ICs rather than glass pre-ICs, where the pre-IC refers to the configuration of simulation particle distribution before adding displacements according to 2LPT.
As shown in Ref.~\cite{masaki21}, the grid pattern of the grid pre-ICs add artificial anisotropy to the simulated matter distribution, especially at high redshifts and on small scales.
We checked that measurements at $z=0$, which we mainly focus on in this work, do not change by the choice of pre-IC. 
Although the observational constraints are tight, e.g., $|\gs|\lesssim {\cal O}(10^{-2})$~\cite{Planck:2018jri, Planck:2019evm, Planck:2019kim, Pullen:2010zy, sugiyama18}, we employ slightly larger values of $\gs=\pm0.1$ and $\pm0.3$ to enhance the signal of 
\Red{the quadrupole power spectra in the simulations, subsequently reducing the noise on the $\bht$ term derived from them.
We confirmed that the results from the runs with $\gs=\pm0.01$ exhibit substantial noise but remain reasonably consistent with those from our main runs shown in the figure presented later.}
We also conduct isotropic simulations with $\gs=0$.
For each value of $\gs$, we run four independent random realizations using the same four random seeds for IC generation, resulting in a total of 20 realizations.

As a sanity check, we first compute the ratio $P_{\rm m,2}(k)/\bigl[\frac{2}{3}\gs\bar P_{\rm m}(k)\bigr]$, where $P_{\rm m,2}(k)$ is the quadrupole moment of the matter power spectrum. 
According to the linear perturbation theory in the anisotropic universe, this ratio should always be unity (see Eq.~\eqref{eq:pksa}).
Therefore, by examining the evolution of this ratio on large scales where linear theory is valid, it should be possible to check whether our simulations are implemented to incorporate SA appropriately.
We measure $\bar{P}_{\rm m}(k)$ from the isotropic simulations of $\gs = 0$.
To cancel out a noise on the ratio, we here employ the ``pairing'' method, which is described below.
Quadrupole power spectra tend to be noisy on large scales due to sample variance.
We found that, roughly speaking, the noise term $\epsilon$ appears on large scales as an additive form: 
\begin{align}
    P_{\rm m,2}(k) \simeq \frac{2}{3}\gs \bar{P}_{\rm m}(k)+\epsilon(k),
\end{align}
where $\epsilon$ fluctuates around zero randomly for individual realizations.
Therefore, by simply averaging the ratios $P_{\rm m,2}(k)/[\frac{2}{3}\gs\bar P_{\rm m}(k)]$ from the two runs with $\gs=\gs^+>0$ and $\gs=\gs^-=-\gs^+$ whose initial random seeds are identical, the noise term is largely canceled.
The results from this paring method are labeled as ``$|\gs|=\gs^+$''.

\begin{figure}
\includegraphics[width=\columnwidth]{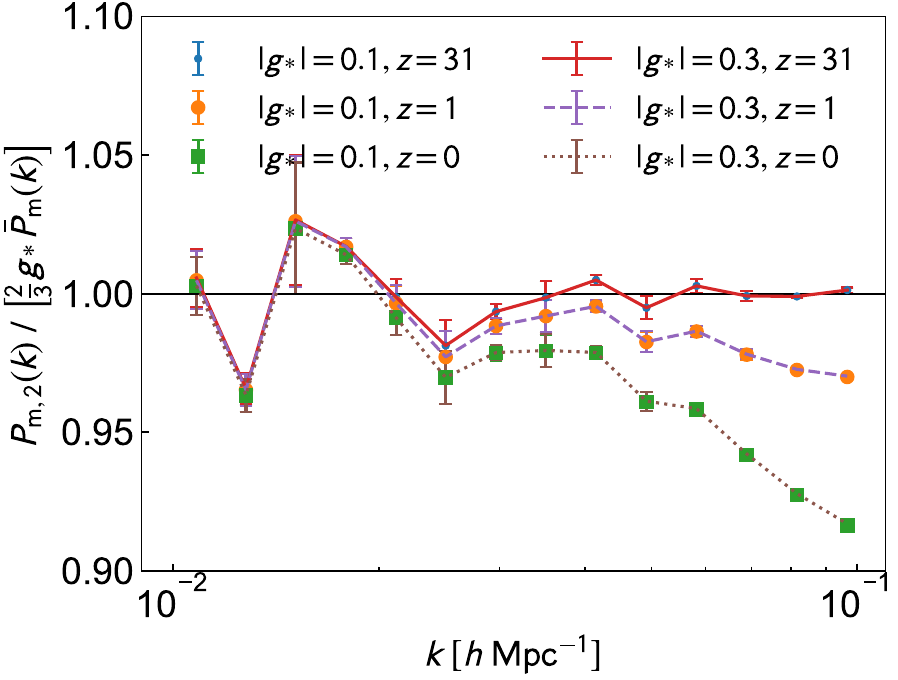}
\caption{The ratio $P_{\rm m,2}(k)/[\frac{2}{3}\gs\bar P_{\rm m}(k)]$ measured from the anisotropic universe simulations at $z=31,~1$ and $0$. The points (lines) with the error bars show the $|\gs|=0.1$ ($0.3$) result. The horizontal solid line shows unity as predicted by the linear theory.}
\label{fig:sanitycheck_pkm}
\end{figure}
In Fig.~\ref{fig:sanitycheck_pkm}, the symbols ($|\gs|=0.1$) and lines ($|\gs|=0.3$) represent the average values of $P_{\rm m,2}(k)/[\frac{2}{3}\gs\bar P_{\rm m}(k)]$ over the four paired realizations, with the standard errors shown as the error bars.
Since we are focusing on linear scales to investigate $\bht$ as predicted by the linear bias model, we plot the data points in the range of $0.01<k/(\hmpc)<0.1$.
The ratio at each redshift ($z=31,~1,$ and $0$) agrees with unity within $5\%$ on larger scales in both cases of $|\gs|=0.1$ and $0.3$, and is independent of the value of $\gs$, 
indicating that the SA has been appropriately incorporated into the simulations.
At later epochs, the ratio slightly decreases toward smaller scales, which may be attributed to nonlinear effects (a detailed study on nonlinear effects will be presented separately; see also Ref.~\cite{Ando08}).

\paragraph{Measurements of $\bht$.---\hspace{-0.8em}}
\begin{figure}
\includegraphics[width=\columnwidth]{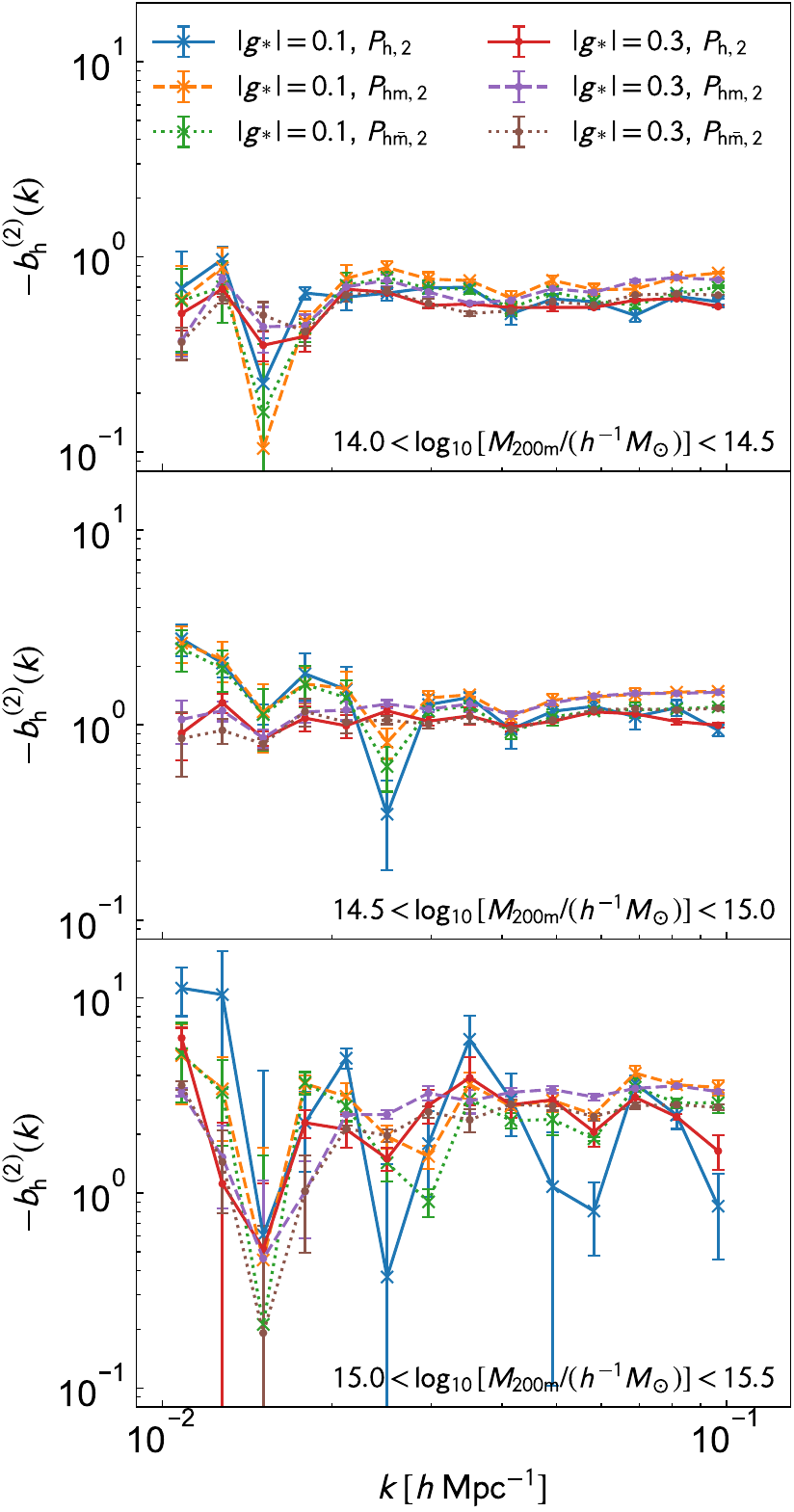}
\caption{The coefficient of the SA-induced bias, $\bht(k)$, measured in our simulations with $|\gs|=0.1$ and $0.3$. The top, middle and bottom panels show the results for the halo mass range of $\log_{10} [M_{\rm 200m}/(\hmsun)] \in (14,~14.5),~(14.5,~15)$ and $(15,~15.5)$, respectively. $\bht(k)$ estimated by Eqs. (\ref{eq:A}), (\ref{eq:B}) and (\ref{eq:C}) are labeled by ``$P_{\rm h,2}$'',~``$P_{\rm hm,2}$'' and ``$P_{\rm h\bar m,2}$'', respectively.
}
\label{fig:bgt}
\end{figure}
To measure $\bht$ from the cosmological $N$-body simulations, we first identify halos using a halo finder {\sc Rockstar} \cite{Behroozi:2013}. 
We use a halo mass definition of $200$ times the mean density denoted as $M_{\rm 200m}$.
We focus on cluster-sized halos with masses in the range of $14.0<\log_{10}[M_{\rm 200m}/(\hmsun)]<15.5$.
Then, to measure $\bht$ in the simulated halo distribution appropriately, we employ three different estimators.
Two of these are defined as
\begin{align}
    \left. \hat{b}_{\rm h}^{(2)}(k) \right|_{\rm h} &=\frac{3 \hat{P}_{{\rm h},2}(k)}{2\gs \bh \hat{\bar{P}}_{\rm m}(k)}-\bh,  \label{eq:A} \\
   \left. \hat{b}_{\rm h}^{(2)}(k) \right|_{\rm hm} &=\frac{3 \hat{P}_{{\rm hm},2}(k)}{\gs \hat{\bar{P}}_{\rm m}(k)}-2\bh,  \label{eq:B}
\end{align}
where quantities with hats denote the values computed from a single realization, and we denote $\hat{b}_{\rm h}^{(2)}(k)$ explicitly to reflect the wavenumber dependence of the estimators.
We precompute $\bh$ in the estimators using the isotropic simulation and simply taking the average value of $\bar P_{\rm hm,0}(k) / \bar P_{\rm m}(k)$ over the four realizations in the range of $0.01<k/(\hmpc)<0.1$.
Here, $\bar P_{\rm hm,0}(k)$ is the monopole component of the cross-power spectrum between the halo density field and the matter distribution in the isotropic simulations.
Thanks to the isotropic simulation which shares the random seed for IC generation with the SA realization, we can compute ${P}_{\rm h\bar m}(\vk)$ -- the cross-power spectrum between the halo density field in the SA simulation $\delta_{\rm h}({\bf k})$ (Eq.~\eqref{eq:hndf}) and the matter distribution in the isotropic simulation $\bar{\delta}_{\rm m}({\bf k})$ (Eq.~\eqref{eq:mdf}) -- in addition to $\bPkm$.
Using these, the third estimator is defined as
\begin{align}
   \left. \hat{b}_{\rm h}^{(2)}(k) \right|_{\rm h\bar{m}} &=\frac{3 \hat{P}_{{\rm h\bar m},2}(k)}{\gs\hat{\bar{P}}_{\rm m}(k)}-\bh\label{eq:C}.
\end{align}
It is straightforward to verify that all the three estimators are constructed to correctly yield the $\bht(k)$ coefficient within linear theory.
For validation under realistic nonlinear conditions, results obtained from these estimators are compared with each other.

Fig.~\ref{fig:bgt} shows the coefficient of the SA-induced bias, $\bht(k)$, measured in our simulations with $|\gs|=0.1$ and $0.3$, using the three estimators.
Note that the figure shows $-\bht(k)$ rather than $\bht(k)$.
The pairing method is also applied here to reduce the sample variance.
As in Fig.~\ref{fig:sanitycheck_pkm}, the symbols and the error bars represent the averages and the standard errors of $\hat{b}_{\rm h}^{(2)}(k)$, respectively, calculated over the four pair realizations.
We consider three halo mass ranges, $\log_{10} [M_{\rm 200m}/(\hmsun)] \in (14,~14.5)$ (top) $,~(14.5,~15)$ (middle) and $(15,~15.5)$ (bottom), and present the results in separate panels.
As shown in this figure, the measured $\bht(k)$ converges to a nearly constant nonzero negative value for the three estimators in the range of $0.01<k/(\hmpc)<0.1$ depending on the halo mass.
The figure also shows the agreement between the results for $|\gs|=0.1$ and $|\gs|=0.3$ at the overall range we consider, and it means that the SA-induced bias in the linear regime should be independent of the SA parameter $\gs$ as expected.
This confirms that the SA-induced bias contribution predicted in the linear model indeed exists for cluster-sized halos on large scales.
Moreover, this paper successfully measured the nonzero value of this bias coefficient for the first time.

\begin{figure}
\includegraphics[width=\columnwidth]{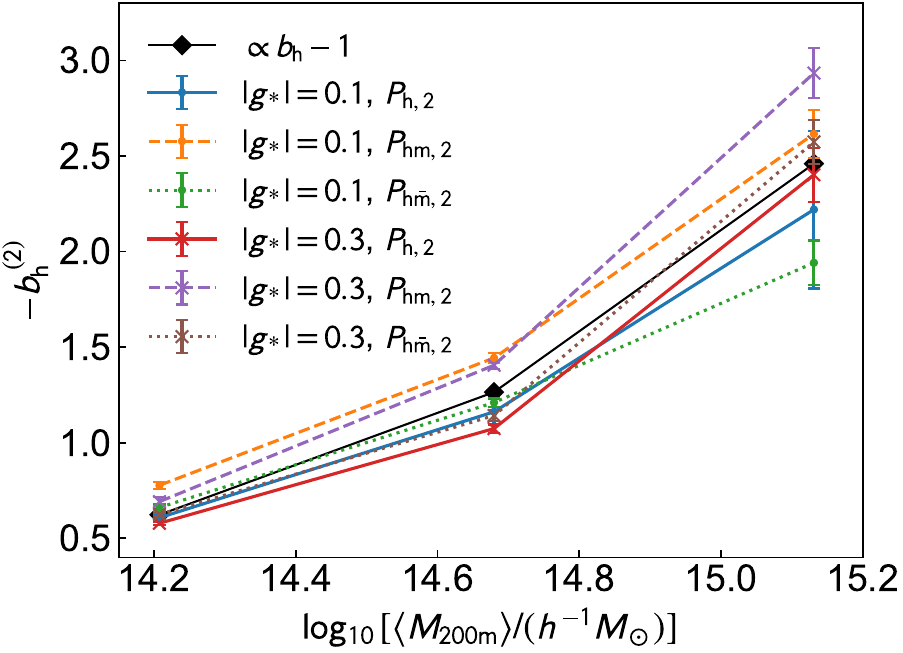}
\caption{The coefficient of the SA-induced bias, $\bht$, obtained from the three estimators (Eqs.~\eqref{eq:A}, \eqref{eq:B} and \eqref{eq:C}) with both $|\gs|=0.3$ and $0.1$ runs as a function of the average halo mass $\langle M_{\rm 200m}\rangle$ in the three mass ranges. 
For comparison, \Redd{$0.45(\bh-1)$} obtained from the isotropic realizations is also shown.
}
\label{fig:bg2_lin}
\end{figure}

To quantitatively assess the halo mass dependence of $\bht(k)$, we fit the measured $\bht(k)$ in the range of $0.01<k/(\hmpc)<0.1$ for each halo mass range shown in Fig.~\ref{fig:bgt} by a constant value in $k$.
The obtained values are denoted as $\bht$, with the $k$-dependence omitted.
Fig. \ref{fig:bg2_lin} shows $-\bht$ obtained from the three estimators (Eqs.~\eqref{eq:A}, \eqref{eq:B} and \eqref{eq:C}) for both the $|\gs|=0.3$ and $0.1$ runs, as a function of the average halo mass $\langle M_{\rm 200m}\rangle$ in the three mass ranges.
We observe a consistent dependence of $\bht$ on halo mass across all six cases, for the different estimators and the different values of $|\gs|$.
For comparison, we overplot \Redd{$0.45(\bh-1)$} obtained from \Redd{our estimates of $\bh$ taking values for $2.4 \leq \bh \leq 6.5$. 
Interestingly, there seems to be a scaling relation, $\bht \propto -(\bh - 1)$, 
as also seen in other-origin tidal biases \cite{baldauf12,akitsu23}.}
As a result, the quadrupole moment is biased differently from the monopole moment, with the ratio being \Redd{$0.62 \lesssim (\bh + \bht)/\bh \lesssim 0.74$}.
In other words, in a statistically anisotropic universe, the quadrupole \Red{cluster-sized} halo power spectra are consistently less amplified than monopoles.

Since our estimates of $\bht$ are derived for the cluster-sized halos, they are applicable to constraining $\gs$ by analyzing the clustering of galaxy clusters \cite{jimeno17}.\footnote{
\Redd{In practical data analyses, non-primordial anisotropies due to, e.g., the redshift-space distortion, the Alcock-Paczy\'nski effect and peculiar survey geometries must be properly subtracted. It would be feasible by means of the polypolar spherical harmonic decomposition technique \cite{Shiraishi:2016wec,shiraishi21,shiraishi23} as demonstrated in Ref.~\cite{sugiyama18}.
}}
\Reddd{
The constraining power from current galaxy cluster surveys would, however, be significantly weak because of their very limited sample numbers and survey volumes. It would be hard indeed to hunt the anisotropic signal with $|\gs| = {\cal O}(10^{-2})$ if any. In light of this, analyzing distributions of (not galaxy clusters but) galaxies is somewhat more promising and could potentially capture it~\cite{Shiraishi:2016wec}. Therefore}, estimating $\bht$ of galaxy-sized halos would also be an important topic.


\paragraph{Conclusions.---\hspace{-0.8em}}
Using cosmological $N$-body simulations incorporating the quadrupolar SA, we studied the coefficient of the SA-induced halo bias, $\bht$, predicted in the linear bias model.
To achieve this, we have introduced three simulation-based estimators defined as Eqs. (\ref{eq:A}), (\ref{eq:B}) and (\ref{eq:C}), and successfully confirmed the existence of the SA-induced halo bias.
We showed that $\bht(k)$ is negative on large scales, approximately constant in wavenumber (Fig. \ref{fig:bgt}), and its absolute value increases with halo mass within the mass range considered in this work (Fig. \ref{fig:bg2_lin}).
As a result, the quadrupole moment of the halo power spectrum gains an additional dependence on the new bias parameter, which is absent for the monopole in a statistically anisotropic universe.
Such anisotropic features in the halo bias provide a new tool for testing the statistically anisotropic universe and are expected to lead to more precise verification of anisotropic inflation scenarios, as well as models involving vector dark matter and dark energy.

In this {\it Letter}, we focused on the halo distribution on large scales, which approximately correspond to the linear regime, and did not address the behavior on smaller scales where the nonlinear structure growth becomes significant.
The nonlinear effects on the SA are intriguing and will be explored in future work.
In addition, we plan to investigate $\bht$ of less massive galaxy-sized halos, its time evolution, and halo bias potentially induced by various forms of the SA, as discussed in, e.g.,  Refs.~\cite{Kehagias:2017cym,Bartolo:2017sbu,Fujita:2018zbr}.
These studies would further provide deeper insights into the SA and its implications for underlying cosmological scenarios.

\paragraph{Acknowledgments.---\hspace{-0.8em}}
\Red{We thank the anonymous referees for their constructive and insightful comments.}
We thank Kazuyuki Akitsu \Red{and Masahiro Takada} for \Red{useful} discussion.
The calculations in part were carried out on Cray XC50 \Red{and XD2000} at Center for Computational Astrophysics, National Astronomical Observatory of Japan.
This work was supported in part by JSPS KAKENHI Grant Numbers JP22K03644 (SM), JP20H05859 and JP23K03390 (MS), 
JP20H05861, JP21H01081, JP22K03634, JP24H00215 and JP24H00221 (TN), 
JP20K03968, JP23H00108 and JP24K00627 (SY).
TO acknowledges support from the Taiwan National Science and Technology Council under Grants No. NSTC 112-2112-M-001-034- and No. NSTC 113-2112-M-001-011- and the Academia Sinica Investigator Project Grant No. AS-IV-114-M03 for the period of 2025–2029.


\bibliographystyle{apsrev4-2}
\bibliography{lssref}

\end{document}